\documentclass[12pt]{iopart}
\usepackage{graphicx}
\usepackage{dcolumn}
\usepackage{color}
\usepackage{bm}
\usepackage{iopams}
\usepackage{iopams}

\def\dof{d.o.f.}
\def\dofs{d.o.f.}
\def\pt{\PD{}{t}}

\newcommand{\aop}{\opr{C}}                      
\newcommand{\amat}{\arrfont{C}}                 
\newcommand{\arrfont}[1]{\mbox{\sffamily$\textbf{#1}$}}
\newcommand\assign{\mathbin{:=}}                
\newcommand{\be}{\begin{equation}}
\newcommand{\bnum}{\lavec{b}}                   
\newcommand{\bvec}[1]{\bf {#1} }                
\newcommand{\conlab}[1]{#1_{\rm{c}}}            
\newcommand{\diag}{{\rm diag}} 
\newcommand{\Dmat}{\arrfont{D}}
\newcommand{\DP}[2]{{#1\boldsymbol{\cdot}#2}}   
\newcommand{\DTmat}{\arrfont{D}^{\rm T}}
\newcommand{\ee}{\end{equation}}
\newcommand{\eq}[1]{(\ref{#1})}
\newcommand{\fig}[1]{figure\ \ref{#1}}
\newcommand{\gaspar}{GASpAR}
\newcommand{\gnum}{\lavec{g}}                   
\newcommand{\grad}{\vec{\nabla}}                
\newcommand{\Hone}{\set{H}^1}                   
\newcommand{\Imat}{\arrfont{I}}
\newcommand{\ipcsd}[3]{\langle#1,#2\rangle_{\rm #3}}                             
\newcommand{\Jmat}{\boldsymbol\Phi}             
\newcommand{\lavec}[1]{{\boldsymbol#1}}         
\newcommand{\Lmat}{\arrfont{L}}                 
\newcommand{\Ltwo}{{\cal L}_2}                   
\newcommand{\mask}{\boldsymbol{\mathsf{\Pi}}}   
\newcommand{\Mmat}{\arrfont{M}}                 
\newcommand{\opr}[1]{\mathcal{#1}}              
\newcommand{\PD}[2]{\partial_{#2}#1}            
\newcommand{\pnum}{\lavec{p}}                   
\newcommand{\ppmnum}{\lavec{p}^{\pm}}           
\newcommand{\polynomialsset}[1]{\set{P}_{#1}}   
\newcommand{\pspace}{\set{Y}}                   

\newcommand{\Qmat}{\arrfont{A}}                 
\newcommand{\qnum}{\lavec{q}}                   
\newcommand{\Rn}{{\sf {R_v}}}
\newcommand{\Sec}[1]{section \ref{#1}}
\newcommand{\set}[1]{\mathbf{#1}}               
\newcommand{\setdef}[2]{\left\{#1\left\bracevert#2\right.\right\}}
\newcommand{\si}{\mu}                           
\newcommand{\trps}[1]{\raisebox{0pt}{$#1$}^{\rm {\sc{T}}}}

\newcommand{\unum}{\lavec{u}}                   
\newcommand{\uspace}{\set{U}}                   
\newcommand{\uv}[1]{{\vec{e}}^{\:#1}}           
\newcommand{\znum}{\lavec{Z}}                   
\newcommand{\zpnum}{\lavec{Z}^{+}}

\newcommand{\zmnum}{\lavec{Z}^{-}}

\newcommand{\zpmnum}{\lavec{Z}^{\pm}}
\newcommand{\zmpnum}{\lavec{Z}^{\mp}}

\def\eg{{\it e.g.}\ } 
 
\def\ie{{\it i.e.,}\ }

\def\u{{\bf u}}  

\def\a{{\vec a}} \def\B{{\vec B}} \def\j{{\vec j}} \def\b{{\vec b}} \def\u{{\vec u}}   \def\Z{{\vec Z}} \def\w{{\vec w}} \def\ztst{{\vec \zeta}}

\begin{document}

\title[Spectral adaptive mesh refinement in MHD]{Adaptive mesh refinement with spectral accuracy for magnetohydrodynamics in two space dimensions}

\author{D Rosenberg$^{1}$, A  Pouquet$^1$ and P D Mininni$^1$}
\address{$^1$ TNT/IMAGe, National Center for Atmospheric Research, P.O. Box 3000,
  Boulder, CO 80307-3000, USA}
\ead{duaner@ucar.edu,  pouquet@ucar.edu, mininni@ucar.edu}

\begin{abstract}
We examine the effect of accuracy of high-order spectral element methods,
with or without adaptive mesh refinement (AMR),
in the context of a classical configuration of magnetic reconnection
in two space dimensions, the so-called Orszag-Tang vortex made up of
a magnetic X-point centered on a stagnation point of the velocity.
A recently developed spectral-element adaptive refinement incompressible
magnetohydrodynamic (MHD) code is applied to simulate this problem. The MHD
solver is  explicit, and uses the Els\"asser formulation on high-order elements.
It automatically takes advantage of the adaptive grid mechanics that have been
described elsewhere in the fluid context
[Rosenberg, Fournier, Fischer, Pouquet, J. Comp. Phys. 215, 59-80 (2006)];
 the code allows both statically refined and
dynamically refined grids. Tests of the algorithm using analytic solutions are
described, and comparisons of the Orszag-Tang solutions with pseudo-spectral 
computations are performed. 
We demonstrate for moderate Reynolds numbers that the algorithms using both static
and refined grids reproduce the pseudo--spectral solutions quite well.
We show that low-order truncation--even with a comparable number of
global degrees of freedom--fails to correctly model some strong (sup--norm) quantities
in this problem, even though it satisfies adequately the weak (integrated)
balance diagnostics.
\end{abstract}

\submitto{\NJP}
\maketitle

\section{Introduction}
\label{Intro}

In geophysical and astrophysical flows, the Reynolds numbers are large, and nonlinear terms appearing in the 
primitive equations lead to strong mode coupling and multiple scale interactions. Moreover, magnetic fields are often 
in quasi-equipartition with the velocity, as is the case in the Solar Wind or in the interstellar medium. However, 
direct numerical simulations (DNS) using regular grids cannot, even to this day, deal with such large Reynolds numbers 
$\Rn$. Doubling the grid resolution (and thus multiplying the Reynolds number by roughly a factor two) comes 
at a cost of increase in needed computer time by a factor of sixteen in three dimensions, assuming the temporal scheme is 
explicit; even when taking Moore's law into account, such an increase in $\Rn$ can only be achieved roughly every six 
years. Thus, one is led to resort to more sophisticated techniques, such as, for example, turbulence modeling 
(see e.g. \cite{meneveau_katz2000}). However, in the case of coupling to a magnetic field, and using the magnetohydrodynamic 
(MHD) approximation valid for the description of the large-scale dynamics, few such techniques have been developed and 
tested (see however \cite{muller,pablo_2d_alpha,pablo_3d_alpha,jon_cancel}). Another venue is to develop
adaptive mesh refinement (AMR) methods. In this context, we examine in this paper the accuracy of an AMR code 
using spectral elements by comparing its output to exact solutions in simplified cases and to computations using 
a pseudo-spectral code at the same Reynolds numbers on a classical configuration of magnetic reconnection in two-dimensional geometry.
 
We set up the equations in the next \Sec{sec_setup},  and in \Sec{sec_code}, 
describe the numerical method for MHD developed within the context of the adaptive spectral element  
method presented in \cite{rosenberg2006} for the two-dimensional Burgers equation. This section also
presents test results for the method. In \Sec{sec_ot} we apply the method to the Orszag--Tang
configuration, and compare with pseudo--spectral results.  
We consider effects of low order versus high order local approximations
in \Sec{sec_high_low_order},
and \Sec{sec_conclusion} is the conclusion, in which we summarize the results, and offer some observations
about the performance of the method and some directions for future work.

\section{Setup and theory}
\label{sec_setup}

\subsection{Equations, code, and simulations}

For an incompressible fluid with constant mass density $\rho_0$, the magnetohydrodynamic  (MHD)
equations read:
\begin{equation}
\partial_t {\u} + {\u}\cdot \nabla {\u} = - \nabla p + \j \times \b
    + \nu \nabla^2 {\u} ,
\label{eq_momentum} \end{equation}
\begin{equation}
\partial_t {\b}  =  \nabla \times (\u \times \b)
    + \eta \nabla^2 {\b}
\label{eq_induction} \end{equation}
\begin{equation}
\nabla \cdot {\u} =0, \hskip0.2truein \nabla \cdot {\b} =0
\label{eq_incompressible}
\end{equation}
where ${\u}$ and $\b$ are the velocity and magnetic field (in Alfv\'en velocity units, 
$\b=\B/\sqrt{\mu_0 \rho_0}$ with $\B$ the induction and $\mu_0$ the permeability); $p$ is the pressure divided by 
the mass density, and $\nu$ and $\eta$ are the kinematic viscosity and the magnetic resistivity.
The mode with the largest 
wavevector in the Fourier transform of ${\u}$ at initial time is $k_0$.
We also define the viscous dissipation wavenumber as 
$k_\nu=(\epsilon/\nu^3)^{1/4}$, where $\epsilon\sim U_0^3/L_0$ is the energy dissipation rate, with $U_0$ the r.m.s. velocity and
$L_0$ the integral length scale (see below). The Kolmogorov scale at which dissipation sets in is defined as $l_D = 2\pi/k_\nu$; the expression for $k_\nu$ is based on a kinetic energy spectrum 
$E_V(k)\sim \varepsilon^{2/3}k^{-5/3}$. A large separation between the two scales ($k_0^{-1} \gg  k_\nu^{-1}$) is required 
for the flow to reach a turbulent state with significant nonlinear interactions.

In the absence of external forcing, viscosity and magnetic resistivity, the MHD 
equations in two space dimensions (2D) conserves the total energy:
\begin{equation}
E = \frac{1}{2} \int{(u^2 + b^2) d {\bf x}^2} \, ,
\label{eq_einvar}
\end{equation}
together with
the cross helicity
$H_c = \frac{1}{2} \int{{\u} \cdot \b  \, d{\bf x}^2} \, ,$
and the $L_2$ norm of the magnetic potential
$M_a = \frac{1}{2} \int{a^2  \, d{\bf x}^2} \, ,$
with $\b=\nabla \times \a$.

The Reynolds number is defined as $\Rn = U_0L_0/\nu$, where the integral length scale of the flow is defined as
\begin{equation}
L_0 = 2\pi \frac{\int{E_V(k) k^{-1} dk}}{\int{E_V(k) dk}} \ .
\label{eq_integral}
\end{equation}
%
The large scale turnover time 
can then be defined as $\tau_{NL}=U_0/L_0$. We can also introduce the Taylor 
based Reynolds number $R_\lambda = U\lambda/\nu$, where the Taylor 
length scale $\lambda$ is given by
\begin{equation}
\lambda = 2\pi \left(\frac{\int{E_V(k) dk}}{\int{E_V(k) k^2 dk}}\right)^{1/2}.
\label{eq_taylor}
\end{equation}
Length scales built with $\b$ and its energy spectrum $E_M(k)$ can also be defined; the magnetic Reynolds number is $R_m = U_0 L_0/\eta$.

\section{Algorithm description for MHD}
\label{sec_code}

For this work, 
we use a spectral element method \cite{patera1984}, encapsulated within the 
Geophysical\ /Astrophysical Spectral--element Adaptive Refinement (\gaspar) code. Aspects of this code--in 
particular those regarding the dynamic grid refinement--have been described elsewhere 
\cite{rosenberg2006}, where
results were presented for the multi-dimensional Burgers (advection--diffusion) equation. 
Here, we extend that development in several distinct ways in order to solve 
\eq{eq_momentum}-\eq{eq_incompressible}, 
specifically, by adding the pressure term and the Lorenz force in the momentum
equation and by taking into account the magnetic induction equation,  and the divergence--free conditions on 
velocity and magnetic field.

We solve equations \eq{eq_momentum}-\eq{eq_incompressible} in Els\"asser form \cite{elsasser1950}:
\be
\partial_t \Z^{\pm} + \Z^{\mp}\cdot\nabla \Z^{\pm} + \nabla p -\nu^{\pm}\nabla^2 \Z^{\pm} - \nu^{\mp} \nabla^2 \Z^{\mp} = 0
\label{eq_zmomentum}
\ee
\be
 \nabla \cdot  \Z^{\pm} = 0\,,
\label{eq_zconstraint}
\ee
where 
$$ 
\Z^{\pm} = \u \pm \b
$$
and
$$ 
\nu^\pm = \frac{1}{2}(\nu \pm \eta).
$$
Thus, we solve for $\u$ and $\b$ in terms of $\Z^{\pm}$. 

Equations \eq{eq_zmomentum}-\eq{eq_zconstraint} can be recast into an equivalent variational 
form on domain $D$ by defining the following function spaces:
\begin{eqnarray}
\uspace_{\vec{b}}&\assign%
\setdef{\w=\sum_{\si=1}^dw^\si\uv{\si}\,}{w^\si\in\Hone({D})\; \forall \mu\;\&\;\w=\vec{w}_0\;{\rm on}\;\partial D}\\
\end{eqnarray}
\begin{eqnarray}
\Hone(D)&\assign\setdef{f\,}{f\in\Ltwo(D)\;\&\;\PD{f}{x^\si}\in\Ltwo(D)\;\forall\mu}r\,,
\end{eqnarray}
where $\w = \u,\, \b$.
Let $\Z^\pm$ 
and $p$ and their test functions, $\ztst^\pm$ and $q$ be restricted to finite--dimensional subspaces of 
these spaces:
\begin{eqnarray}
\Z\pm     \in \uspace^{N}     = \uspace_{\vec{b}}  &\; \bigcap &\; \polynomialsset{N}\\
\ztst^\pm \in \uspace_{0}^{N} = \uspace_{\bvec{0}} &\; \bigcap &\; \polynomialsset{N}\\
p, q      \in \pspace^{N-2}   = D           &\; \bigcap &\; \polynomialsset{N-2}.
\end{eqnarray}

The basis for the velocity expansion in $\polynomialsset{N}$ is the set of 
Lagrange interpolating polynomials on the Gauss-Lobatto-Legendre ({\rm GL}) quadrature nodes, and 
the basis for the pressure is the set of Lagrange interpolants on the 
Gauss-Legendre ({\rm G}) quadrature nodes. 
Then, the equations \eq{eq_zmomentum}-\eq{eq_zconstraint} can be written in weak form as
\cite{fischer1997}

\begin{equation}
\ipcsd{\ztst^\pm}{\pt \Z^\pm}{GL}+\ipcsd{\ztst^\pm}{\aop^\mp \Z^\pm}{GL} -\frac{1}{\rho_0}\ipcsd{ p}{\nabla\cdot\ztst^\pm}{G}  \\
=-\nu^\pm\ipcsd{{\smash{\vec{\vec{\nabla{\zeta^{\pm,{\rm T}}}}}}}}{\vec{\vec{\nabla{Z^\pm}}}}{GL} 
\label{eq_zquadraturemom} 
\end{equation}
\be
\ipcsd{q}{\nabla\cdot \Z^\pm}{G}  = 0,
\label{eq_zquadraturediv} 
\ee
where $\aop^\pm\assign\DP{\Z^\pm}{\grad}$ is the continuous advection operator, $\ipcsd{\cdot}{\cdot}{GL}$, represents the inner product using quadrature on the {\rm GL} nodes, 
and $\ipcsd{\cdot}{\cdot}{G}$ indicates inner product using quadrature on the {\rm G} nodes.
Thus, we use a {\it staggered grid}, where the quantities ($\u$, $\b$, $\Z^\pm$), on the
{\rm GL} nodes are continuous, while those on the {\rm G} nodes ($p$) are not. This 
so-called $\polynomialsset{N}-\polynomialsset{N-2}$ was chosen to prevent spurious pressure modes 
\cite{maday1992,fischer1997}.

In the spectral element method, functions in $\uspace^{N}$ and $\pspace^{N-2}$ are represented 
as expansions in terms of tensor products of basis functions within each subdomain, or element, the non-overlapping
union of which comprises the domain \cite{patera1984}: 
$ D = \bigcup_{k=1}^{K} \set{E}_k$.

By expanding $\Z^\pm$ and $p$ in terms of their basis functions, inserting these expansions into
\eq{eq_zquadraturemom}-\eq{eq_zquadraturediv},
and using the appropriate quadrature rules,
we arrive at a set of semi-discrete equations in terms of spectral element operators:
\begin{eqnarray}
\Mmat \frac{d\zpmnum_j}{dt} &= -\Mmat\amat^{\mp}\zpmnum_j + \DTmat_j \ppmnum -\nu_{\pm}\Lmat\zpmnum_j -\nu_{\mp}\Lmat\zmpnum_j 
\label{eq_zpsemidiscrete}\\
\Dmat^j \zpmnum_j & = 0,
\label{eq_divsemidiscrete}
\end{eqnarray}
for the $j^{\rm th}$ components of momentum. In this equation, $\Mmat$, $\Lmat$, and $\amat$, are the well--known
mass matrix, weak Laplacian and advection operators, respectively (\eg, \cite{rosenberg2006}, and references
therein). 
The variables $\vec{\znum}^\pm$ represent values of the $\Z^\pm$ collocated at the GL node points, while 
$\pnum^\pm$ are values of the pressures at the G node points. Similarly, we denote the collocated 
test functions by $\qnum$.
The {\it Stokes operators}, $\Dmat_j$, arise from the quadrature rule in \eq{eq_zquadraturediv}, in which
the {\rm GL} basis function and its derivative must be interpolated to the {\rm G} node points, and multiplied by the
{\rm G} quadrature weights. In two dimensions, 
\be
\ipcsd{\qnum}{\nabla\cdot\vec{\znum}}{G} = \sum_{k=1}^{K} \trps{(\qnum^k)} ( \Dmat_{1}^{k} \znum_{1}^{k}
+\Dmat_{2}^{k} \znum_{2}^{k}) 
\ee
for $\vec{\znum}^k=\vec{\znum}^{\pm,k}$.
For regular rectangular elements, of lengths $L_{j}^{k}$,
$$
\Dmat_{1}^{k} = \left(\frac{L_{2}^{k}}{2}\right) \tilde{\Imat} \otimes \tilde{\Dmat}\, , \quad
\Dmat_{2}^{k} = \left(\frac{L_{1}^{k}}{2}\right) \tilde{\Dmat} \otimes \tilde{\Imat},
$$
where
$$
  \tilde{\Imat}_{ij} = \sigma_i \phi_{j}(\eta_i), 
$$
and
$$
  \tilde{\Dmat}_{ij} = \sigma_i \frac{d\phi_{j}}{dr}|_{r=\eta_i} 
$$
are, respectively, the weighted one-dimensional (1D) interpolation matrix mapping {\rm GL} points to the {\rm G} points,
and the weighted 1D {\rm GL} derivative matrix interpolated to the {\rm G} points \cite{fischer1997}.
In these definitions $\sigma_i$ are the weights corresponding to the G node points.
Just like with the mass and Laplacian operators, the Stokes operators are
written as tensor products of 1D operators.

The above equations, \eq{eq_zpsemidiscrete}-\eq{eq_divsemidiscrete},
are correct for a single element, but they are not complete when we
have multiple subdomains. In this case, we must ensure that all quantities in $\uspace^{N}$ 
remain continuous across element interfaces. The manner in which this is
done for advection--diffusion on (non)conforming elements was described in
\cite{rosenberg2006}. Using the Boolean scatter matrix, $\conlab{\Qmat}$, the interpolation
matrix from global to local degrees of freedom, $\Jmat$,  and the masking matrix (that 
enforces homogeneous boundary conditions), $\mask$, that were presented there,
we find that we must replace the local Stokes operators above with 
$$
\Dmat_j \to \Dmat_{L, j} \Jmat\conlab{\Qmat}\mask, 
$$
where $\Dmat_{L, j}=\diag_k(\Dmat_{j}^{k})$, and the $\Dmat_{j}^{k}$ are the
matrices from above. In much of what follows, we will continue to use the local 
form of the Stokes operators, and 
simply recognize that the multiple subdomain form can be imposed.

Note in \eq{eq_zpsemidiscrete} the presence of different 
pressures for $\Z^\pm$. As we show below,
we will maintain the divergence constraints by solving for the pressures for both $\Z^\pm$. While
analytically these pressures are the same, they serve as Lagrange multipliers
\cite{ferziger2002} for their respective fields, $\Z^\pm$. Given that each field has its own
constraint that is solved independently of the other, numerically they will in general be different. 

\subsection{Time stepping}

Because we want to resolve all time scales (as well as spatial scales), we choose
an explicit method for integrating \eq{eq_zpsemidiscrete}-\eq{eq_divsemidiscrete} in time. The method we 
use is the $m^{th}$--order Runge--Kutta (RK), and since the right--hand side of the equations
has no explicit dependence on time, we can use the following formulation \cite[p. 109]{canuto1988} in order to solve $dU/dt = F(U)$:
\begin{center}
\begin{tabular}{ll}
{Set}\\
\hspace{0.5in} $U = U^n$\\
{For $k = m, \; 1, \; -1$ }\\
\hspace{0.5in} $U = U^n + \frac{1}{k}\Delta t F(U)$\\
{End for}\\
\hspace{0.5in} $U^{n+1} = U$.
\end{tabular}
\end{center}

Considering one component of the Els\"asser variables
and following the above RK algorithm, we write for each iteration 
(recall \eq{eq_zpsemidiscrete})
\be
\znum_{j}^\pm = \znum_{j}^{\pm,n}  - \Mmat^{-1}(\Mmat\amat^{\mp} \znum_j^\pm  -  \DTmat_j \pnum^\pm +\nu_{\pm}\Lmat\znum_j^\pm +\nu_{\mp}\Lmat\znum_j^\mp ) .
\label{eq_rkstage}
\ee
We insist that each RK stage obey \eq{eq_divsemidiscrete} in its discrete form, so multiplying
\eq{eq_rkstage} by $\Dmat_j$, summing, and setting the term $\Dmat^j \znum_j^\pm = 0$ leads to the 
following pseudo-Poisson equation for the pressures, $\pnum^\pm$:
\be
\Dmat^j \Mmat^{-1} \DTmat_j \pnum^\pm = \Dmat^j \gnum_j^\pm,
\label{eq_pseudopoisson}
\ee
where the remaining inhomogeneity
$$
\gnum_j^\pm =  \Dmat_j \Mmat^{-1}(\Mmat \amat^{\mp}\znum_j^\pm  + \nu_{\pm}\Lmat\znum_j^\pm +\nu_{\mp}\Lmat\znum_j^\mp ) - \Dmat_j \znum_j^\pm.   
$$
The pseudo--Laplacian operator on the left--hand--side of \eq{eq_pseudopoisson} also arises from 
a second order implicit time descretization of the spatially discretized equations 
\eq{eq_zpsemidiscrete}-\eq{eq_divsemidiscrete} 
as shown in \cite{fischer1997,kruse2002}.
In this formulation, even if $\znum_j^\pm$ is not divergence--free, the partial update after this RK stage will be.
The inverse mass operator, $\Mmat^{-1}$, must be computed for a given grid configuration. For conforming
elements, this matrix is trivially inverted since it is diagonal. For nonconforming discretizations, 
the mass matrix is lumped in order to recover a diagonal matrix that can be
inverted straightforwardly \cite{kruse2002}. Equation \eq{eq_pseudopoisson} is solved using
a preconditioned conjugate gradient method (PCG; see \cite{shewchuck94,WeisLQ2}, and also \cite{rosenberg2006}
for modifications required for nonconforming elements). For preconditioning,
we use a block Jacobi preconditioner computed using either a fast diagonalization method 
\cite{DFM2002}, or a direct inversion. 

Thus, at each timestep, $m$ RK stages are computed, and each stage solves \eq{eq_pseudopoisson} twice, 
once for $\zpnum$, and once for $\zmnum$ leading to $2m$ pseudo-Poisson solves at each time step.
The MHD solver is encapsulated within a class as are all solvers in the code. 

Currently the solver considers the Els\"asser variables, $\znum^{\pm}$, to be auxillary variables, 
so a transformation is done from the native $\unum$, $\bnum$, and back.  However, it is easy to create
a new constructor for the solver object, so that $\znum^{\pm}$ are the primary
quantities of interest. In all results presented in this work, we choose $m=2$ for the RK order.

\subsection{Tests with exact solutions}

To test the algorithm described above, we first consider two test problems that have
analytic solutions. The first case tests the algorithm without magnetic field ($\b\equiv 0$)
field--\ie solves the Navier--Stokes equations--and the second case tests the full MHD algorithm. 

\begin{figure}
\begin{center}
\includegraphics[width=.38\textwidth]{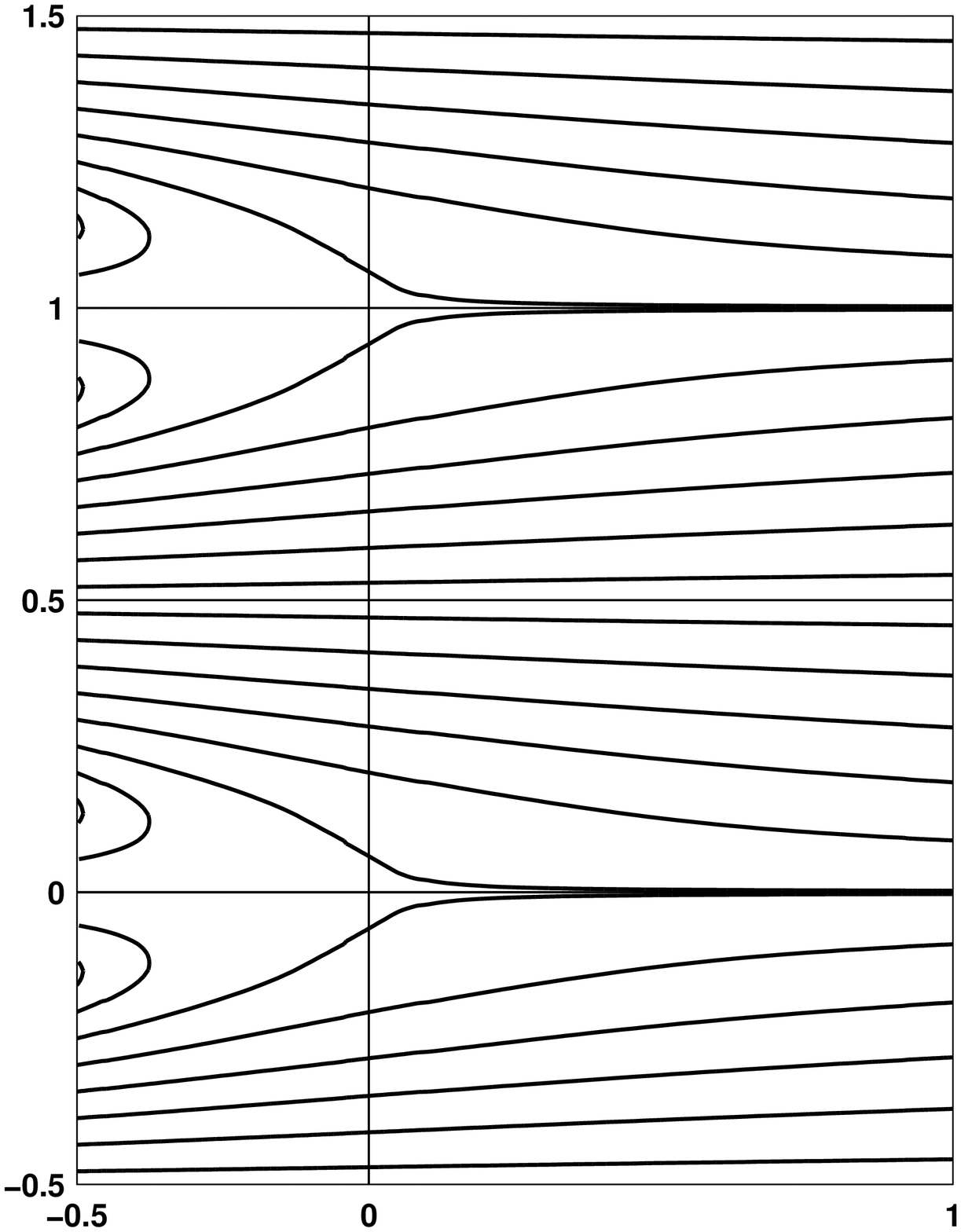}
\includegraphics[width=.56\textwidth]{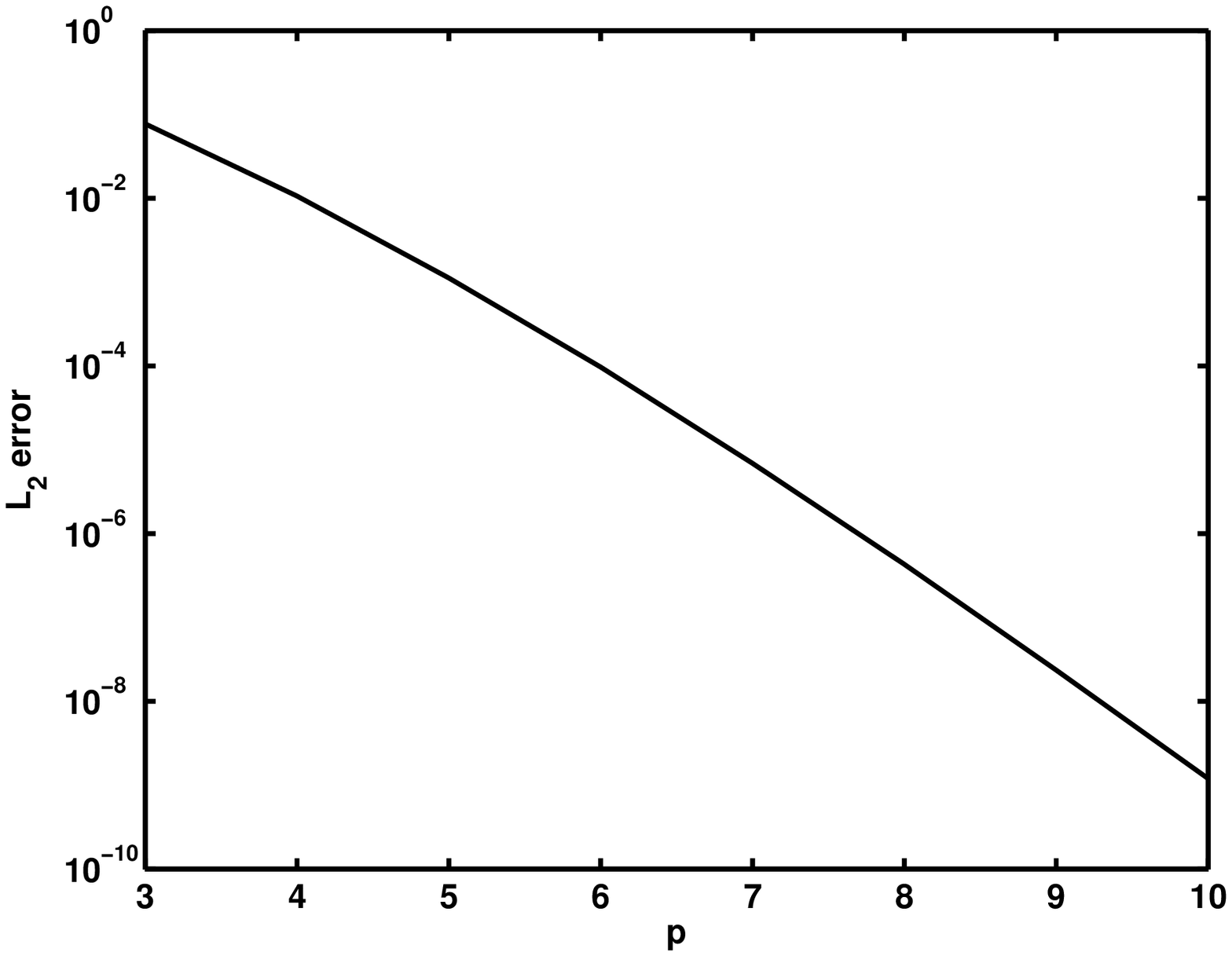}
\end{center}
\caption{({\it left}) Elemental decomposition and streamlines of the steady--state Kovasznay 
solution as determined by time marching for a case where $p=6$. ({\it right}) $\Ltwo$ 
error vs. polynomial degree, showing spectral convergence of the Kovasznay solution.}
\label{fig_kovconv}
\end{figure}

For Navier--Stokes, we use a steady state solution of a (Kovasznay) flow field behind a periodic
array of cylinders \cite{kovasznay1948} (see also \cite{DFM2002}):
\begin{eqnarray*}
u_x &= 1-e^{\lambda x} \cos 2\pi y \\
u_y &= \frac{\lambda}{2\pi} e^{\lambda x} \sin 2\pi y,
\end{eqnarray*}
where $\lambda = \frac{\Rn}{2} -\sqrt{\frac{\Rn^2}{4} + 4\pi^2}$.
We initialize the grid with the solution, and march to a steady state (typically
to $t\approx2$), where we compare the numerical solution with the analytic solution. 
For the following test, the grid we use is conforming, and non-adaptive (see \fig{fig_kovconv});
the time step is fixed at $1\times 10^{-3}$, while 
$\Rn = 1/\nu = 40$. Dirichlet boundary conditions are set from the analytic solution.
We plot the $\Ltwo$ norm of the error
as a function of polynomial degree, $p$ in \fig{fig_kovconv} (right), in order to show convergence.
Clearly, the solution is converging spectrally as desired.  

The next test looks at a steady Hartmann flow,
consisting of a flow of viscous conducting fluid between two parallel plates, with 
separation $2a$, and with a 
magnetic field, $B_0$, applied in a direction perpendicular to the plates. The solution
is \cite{landau1984}:
$$
u_x = u_0\frac{\cosh H_a - \cosh(y H_a/a)}{\cosh H_a - 1} 
$$
$$
b_x = -u_0\sqrt{\frac{\nu}{\eta}}\ \frac{ y/a \, \sinh H_a - \sinh( y H_a /a) }{\cosh H_a - 1}
$$
where the {\it Hartmann number}, 
$$
H_a = \frac{a}{\delta} = \frac{a B_0}{\sqrt{\nu\eta}},
$$
and $\delta$ represents a boundary layer thickness over which the velocity
compared with that in the central region decreases at the top and bottom plates.

Again, we initialize a grid of  $K=4\times2$-elements--whose lower left and upper right 
global grid vertices are $(0,0)$, and $(4,2)$--with the analytic solution, at fixed $p$  and march to 
steady--state, comparing the numerical an analytic
solutions. For these runs we use a Courant--limited timestep. Dirichlet boundary conditions are
again set from the analytic solution.
To drive the flow, we add a constant $x$-momentum forcing, $\lavec{\gamma}_x$, to the 
right--hand--side of \eq{eq_momentum} (or, equivalently, to \eq{eq_zmomentum}) such that
$$
\lavec{\gamma}_x = \frac{u_0^2\ H_a}{a\ \Rn} \frac{\sinh H_a}{\cosh H_a - 1} \ .
$$

\begin{figure}
\begin{center}\includegraphics[width=.52\textwidth]{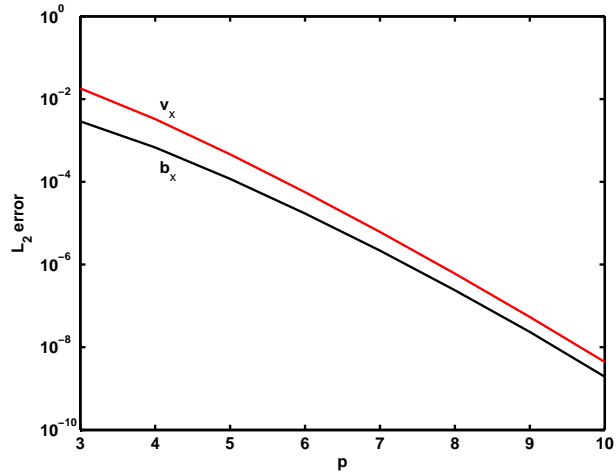}\end{center}
\caption{$\Ltwo$ error vs. polynomial degree $p$, showing spectral convergence of 
the Hartmann solution for $u_x$ and $b_x$. Similar convergence is seen for $u_y$.
}
\label{fig_hartconv}
\end{figure}

For all our tests, we set $H_a=4$, $\Rn \equiv \frac{u_0 a}{\nu} = 40$, $u_0=1$, and $B_0=1$; the choice
of $H_a>1$ suggests a flow where the magnetic field is dynamically significant.
In \fig{fig_hartconv} we present our convergence results. As expected, we again see
spectral convergence.

\section{The Orszag-Tang vortex}
\label{sec_ot}
The Orszag-Tang vortex (hereafter, OT; \cite{OT1979}) is a simple configuration with a magnetic X-point centered at a stagnation point of the velocity. It is best expressed in terms of the stream function $\Psi$ (with $\u=\nabla \times \Psi\,\hat{\lavec{z}} $) and the magnetic potential. It reads
\begin{eqnarray*}
\Psi & = 2 \ \alpha \ [ \cos(2\pi x) + \cos(2\pi y) ] \\
a_z &= 2\cos(2\pi x) + \cos(2\pi y).
\end{eqnarray*}

The total energy $E_T=E_V+E_M$, for $\alpha=1$, is equally divided initially between its kinetic and magnetic components $E_V$ and $E_M$, both equal to 2, and the initial correlation coefficient ${\tilde {\rho}}_c$ is equal to 
$41\%$, where
${\tilde {\rho}}_c={{2\u \cdot \b}\over{u^2+b^2}}$. This configuration is known to develop several current sheets in a time of order unity in these units; such current sheets further destabilize in time through a tearing mode instability embedded within a turbulent flow and develop numerous complex small-scale structures (see e.g.\cite{politano89}). 

In this section, we apply the algorithm described in 
Section 3 to the OT problem.
For adaptive runs, a spectral estimator refinement criterion
is used \cite{henderson99,mavriplis94,rosenberg2006} that estimates the solution error. 
If, in a given element, 
this error is greater than a specified tolerance, $\varepsilon_{\rm est}$, the element is tagged
for refinement. The nominal resolution of the adaptive runs is given by an {\it equivalent} resolution, which
is computed by 
$$
N_{\rm eq} = p N_0 2^{\ell_{\rm{max}}},
$$
where $N_0$ is the initial number of elements in either direction (the {\it base} grid), 
and $\ell_{\rm{max}}$ is the maximum refinement
level. All computations are performed on a periodic grid of dimension $[0,1]^2$. We compare the
spectral element solutions with those obtained from a well--characterized pseudo--spectral code
that has been used to produce numerous results cited in the literature \cite{gomez2005a, gomez2005b}.

\begin{figure}
\begin{center} 
\includegraphics[width=8.3cm]{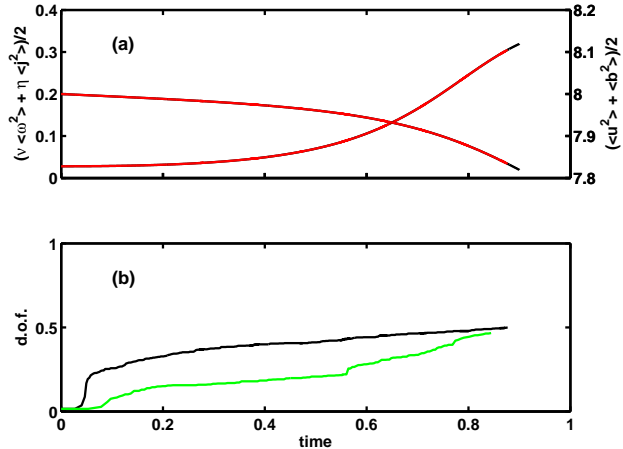}
\end{center}
\caption{(a) Comparison of enstrophy (red curve) with a pseudo-spectral code with $512^2$ \dof\ 
(black curve), as a function of time. The curves overlie each other. (b) Degrees of freedom in the 
adaptive run normalized by the total number of \dof\ in the pseudo-spectral run, as a function of time,
for the refinement criteria based on $u_x$ and $B_x$ (black), and $j$ and $\omega$ (green).
}
\label{comp_enst_dof}
\end{figure}

The total dissipation in the flow is defined as
\be
{\cal D}_T=-\nu < \omega^2> - \eta <j^2>,
\label{eq_diss}
\ee
where $\omega=\nabla \times \u$ is the vorticity and $\j = \nabla \times \b$ is the current density.
It is a global quantity, as is the total energy $E_T$, and is characteristic of the dynamical evolution 
of the flow as a whole: as small-scale gradients in both the velocity and the magnetic field develop 
through non-linear interactions, current and vorticity sheets form and dissipation sets in at 
a time of order unity. 
The temporal evolution of $E_T$ and ${\cal D}_T$ are shown in \fig{comp_enst_dof} for the 
\gaspar\ MHD run (solid line) and the pseudo-spectral code (dotted line) for a fiducial run with 
$\Rn=10^3\pi$; the initial grid is $K=8\times 8$ with $\ell_{\rm{max}}=3$ and $p=8$
(implying $N_{\rm eq} = 512$),
and $\varepsilon_{\rm est} = 1\times 10^{-5}$. The adaptive code 
reproduces the temporal characteristic times 
(as well as secondary maxima in ${\cal D}_T$ at lower Reynolds number, shown in figure \ref{fig_bal_p} 
in the context of a fixed grid, see below); it also reproduces
the amplitude of the global phenomenon of reconnection of 
magnetic field lines and ensuing dissipation of energy. 

The total number of degrees of freedom (\dofs) normalized by the number of modes in the pseudo-spectral run 
used in this comparison during the temporal evolution of the flow is also shown in \fig{comp_enst_dof} in black.
The \dofs\ quickly increases because the initial grid was coarse ($K=8\times8$ elements,
at $p=8$). After this, the \dofs\ grows regularly 
in anticipation of the peak in total dissipation at a slightly later time; note that it is about one-half that of the pseudo-spectral computation.
Other runs at lower Reynolds numbers indicate that,
 beyond this peak, the number of \dofs\  is roughly constant until about $t=1.8$, where the grid appears to anticipate a secondary peak in ${\cal D}_T$ that begins at about $t=2.5$, and corresponds to renewed reconnection of current layers
\cite{politano89} (not shown). 

\begin{figure}
\centerline{\includegraphics[width=8.3cm]{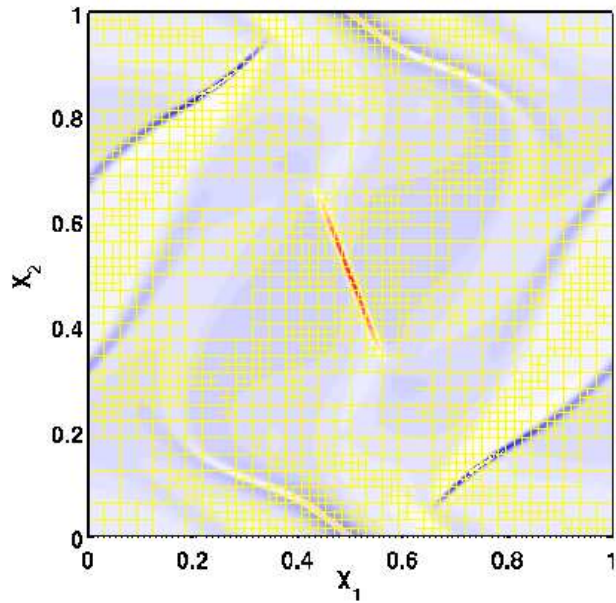}}
\caption{Contours of current density, $j$, for the fiducial run, showing the refined grid. Each yellow box 
is an element; the grid lines indicating the node points within each element are not shown.}
\label{j_contours}
\end{figure}

As the relative number of \dof\  increases toward unity, the AMR becomes less efficient. In this run, the grid refines on the native solver variables $\u$ and $\b$, and by the time the
nonlinear regime begins, the small scales dominate
the flow requiring finer elements to remain on the grid to resolve them. The number of \dofs\ also depends on the criterion of refinement and on $\varepsilon_{\rm est}$; note that here $\varepsilon_{\rm est}$ is set 
tightly so that the grid is more likely to refine. It will be left to 
future work to investigate this and other refinement criteria more systematically, in order 
to determine what range of parameters, and, indeed, which criteria, are the most robust and
efficient for adaptive modeling of flows such as OT. We have freedom in the code also to define
new variables to which to apply our refinement criteria, providing yet another avenue
for investigation (see also the discussion below around \fig{j_contours}).
As an illustration, we also show in \fig{comp_enst_dof} the number of \dofs\ for a spectral criterion 
of refinement, based this time on the vorticity and the current. In this case, the refinement of the 
grid starts at a later time, once the strong gradients have developed, and thus the run is roughly twice 
as fast; however, once we approach the saturation of the growth of small scale production, the number 
of \dofs\ for the two refinement criteria are comparable.

Conservation of energy (and that of the other quadratic invariants) is an essential feature in the 
detailed dynamical evolution of a turbulent flow; it is at the foundation of the concept of energy 
(or invariant) cascade and leads, assuming a constant energy dissipation rate $\epsilon$ within the 
cascade, to the celebrated Kolmogorov energy spectrum $E_V(k)\sim k^{-5/3}$. Note that in MHD, the 
power law followed by the energy spectrum in the inertial range is less clear, and other spectra can 
be postulated a priori on the basis of Alfv\'en wave propagation, including an anisotropic component 
of the spectrum linked with the bi-dimensionalization of the flow in the presence of a strong uniform 
magnetic field. Such power laws are barely observable, due to a variety of reasons. In the laboratory 
or in observations of geophysical flows, the instrumentation has cut-off frequencies, and in numerical 
simulations, the resolution is barely sufficient to be able to resolve the ranges necessary to 
follow the evolution of the flow, namely the energy-containing range around $k\sim k_0$, the 
inertial range where loss-less transfer of energy occurs, and the dissipation range. For example, 
for fluids in three dimensions, it was shown in 
\cite{alex_long} that the Kolmogorov energy spectrum, on resolutions of up to $1024^3$ points, occurs on a small range of wavenumbers (slightly more than two octaves) and is followed by a shallower power law, named 
a bottleneck effect. 

\begin{figure}
\begin{center}\includegraphics[width=8.3cm,angle=-90]{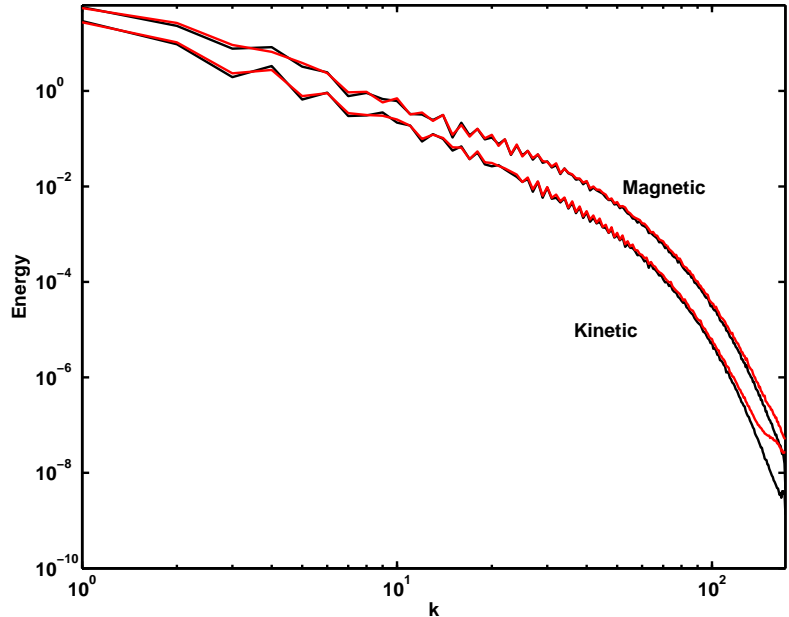}
\end{center}
\caption{ Energy spectra for the fiducial run at a time near the peak in enstrophy.
 The black curves represent the spectra computed
from the pseudo-spectral method, and the red, from \gaspar. The upper curves are the magnetic
energy, while the lower are the kinetic energy.}
\label{spect_512}
\end{figure}

The energy spectra  close to the maximum of enstrophy, at $t\approx 0.85$, are given in 
\fig{spect_512}; the spectra for the AMR run are computed on an irregular grid using the algorithm derived in \cite{fournier2006}. We see that the agreement is quite good between the adaptive 
spectral element run and the pseudo-spectral run. It is interesting to note that, while the 
pseudo-spectral case uses dealiasing, no explicit dealiasing is required in the
\gaspar\ run.

\begin{figure}
\centerline{\includegraphics[width=8.3cm]{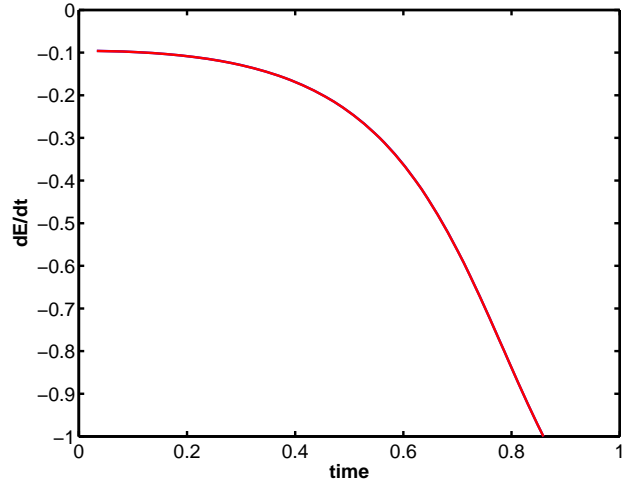}}
\caption{Time variation of the energy invariant 
compared to its exact dissipation when $\nu\not= 0$, $\eta\not= 0$. The blue curve is the time derivative and the red, the
dissipation term. Both curves are normalized by the maximum in the dissipation. There is no discernible difference between
the two curves.  
}
\label{dedt}
\end{figure}

A further test of the code is to verify that the nonlinear terms of the primitive equations do 
preserve the invariants, 
as spectral methods are expected to be conservative. This can be done 
by computing the time derivative of the invariant (say, using an algorithm of order $2$), and
comparing it to its theoretical value.
In the case of the total energy, for example, the 
dissipation ${\cal D}_T$ given in equation \eq{eq_diss} is the theoretical value.
In \fig{dedt} we show the degree to which the energy is preserved and observe again
 an excellent agreement. 

A snapshot of the current density shown in \fig{j_contours} near the peak
in enstrophy indicates that, as expected, the grid refines in the region of the current sheets (and 
vorticity sheets, which are known to be almost co-located). However,
when compared with runs from other authors \cite{friedel1997} using finite difference methods and configuration space 
refinement on the gradients of the basic variables, there appears in our run
to be more refinement than is initially expected outside of the sheets. In the case of the spectral refinement criterion applied
to $\u$ and $\b$, which are native to the solver, the criterion appears to 
capture the variation in the curvature of $\u$ and $\b$ near the current sheets to an extent that may not be required 
since, with the same $\varepsilon_{est}$, half the \dofs\  is needed at early times with the spectral criterion based 
on $j$ and $\omega$.

\begin{figure}
\centerline{\includegraphics[width=8.3cm]{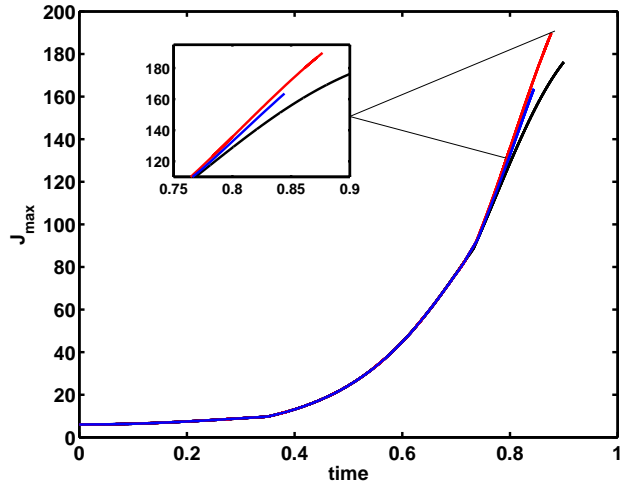}}
\caption{Plot of $J_{\rm max}$ for the fiducial run with the refinement criteria on $u_x$ and $b_x$ (red curve), and on $j$ and $\omega$ (blue). The
black curve is the pseudo--spectral run. 
}
\label{jmax512}  \end{figure}

We now examine the temporal evolution of the current maximum in this fiducial run, as shown in figure \ref{jmax512}. 
As is well known, the development of singularities in such flows may be diagnosed by a $1/[t_*-t]$ behavior \cite{singu_MHD},
with $t_*$ the time of singularity. As such, the temporal evolution of extrema is an important feature of turbulent flows; from 
a physical point of view, such extreme events are also the locus of anomalous diffusion and concentration of (say, chemical) 
tracers so that a faithful reproduction of such events may be required. 
The linear (exponential) growth of the current instability is well reproduced but as we enter the nonlinear phase, 
studied in the literature in the context of the nonlinear growth of the tearing mode instability \cite{rutherford}, errors appear 
when using the AMR codes, errors that are comparable for the two refinement criteria just described (see \fig{jmax512}). As already shown in three 
dimensions \cite{m1536}, the nonlinear phase is highly nonlocal, which indicates that many scales are interacting in the development 
of current instabilities. This discrepancy (not observable in the ${\cal L}_2$ norm) might be remedied by tightening the refinement criteria.
However, the accuracy of the code, as measured primarily by the polynomial order in each element for a fixed number of 
elements and without refining the grid, is a parameter that must also be examined, and this is done in the next section.

\section{High versus low order}
\label{sec_high_low_order}
We now consider the behavior of the OT solutions when the number of global
degrees of freedom is kept (roughly) constant, while the polynomial truncation (degree)
varies. In the first series of runs, we have $\nu\, =\, \eta= 0.025$ ($\Rn = 80\pi$), and these runs are compared with a pseudo--spectral run
with resolution of $128^2$ grid points. 
Here, as stated before, we use a static conforming mesh for each run so
that the results will not be affected by refinement and coarsening
criteria, since we want to focus on the effect the order of the method has on the results. In table 
\ref{tbl_runs}, we present the relevant run parameters.
\begin{table} 
\caption{Parameters used in the simulations described in Section 5; $p$ is the polynomial order in each direction for
         each element, $N_{\rm eq}$ is the linear grid resolution, and $E$ is the number of elements
         in each coordinate direction such that the total number of elements is $K = E \times E$.
         }
\label{tbl_runs}
\begin{center}
\begin{tabular}{c|ccc||cccc}
\hline
             & $p=3$ & $p=4$ & $p=5$ & $p=3$ & $p=4$ & $p=6$ & $p=8$ \\
\hline
$N_{\rm eq}$ & 129   & 128   & 130   & 258   & 256   & 258   & 256    \\
$E$          & 43    &  32   & 26    & 86    & 64    & 43    & 32     \\
\end{tabular} 
\end{center} 
\end{table}

\begin{figure}
\begin{center}\includegraphics[width=8.3cm]{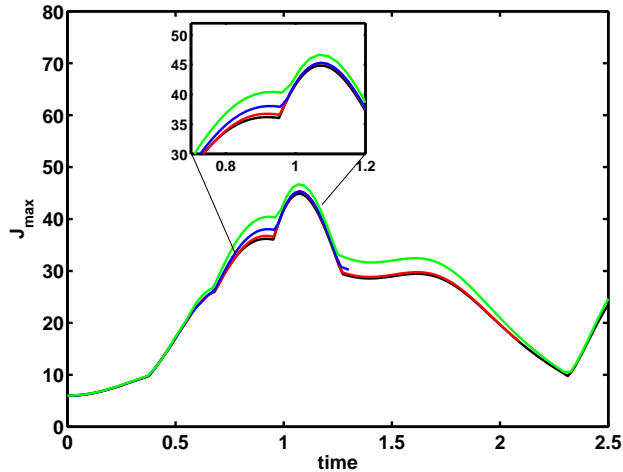}
\end{center}
\caption{Maximum of current $J_{\rm max}$ as a function of time for the first series of $N_{\rm eq} \approx 128$ runs in Table 
\ref{tbl_runs}. The black curve
is the pseudo--spectral case; red is the $p=5$ run; dark blue is the $p=4$ run; green is the
$p=3$ case. Note that the $p=5$ and pseudo--spectral results are
nearly coincident. }  
\label{fig_jmax_128_p}
\end{figure}

In \fig{fig_jmax_128_p} is shown the profile of $J_{\rm max}$ as a function of time for
each of the $N_{\rm eq}\approx 128$ runs. It is seen immediately that the low order truncation does not yield
accurate values for $J_{\rm max}$, particularly as the peak in total dissipation is reached 
near $t=1.1$. 
Clearly, $J_{\rm max}$ converges to the correct
solution as $p$ increases (compare the red curve for $p=5$ and the black curve for the pseudo-spectral run). 
The maximum of vorticity behaves in much the same way as $J_{\rm max}$. 

At this stage, it is desirable to compare the accuracy for a pseudo-spectral code with $N$ points per linear dimension, 
$\epsilon_{ps}$, to the error, $\epsilon_{se}$, of a spectral element code with $E$ elements per linear dimension, each with polynomials of order $p$.
Omitting prefactors with slower variations in the truncation orders, we have for the pseudo--spectral error \cite[p. 400]{canuto1988} that 
$$
\epsilon_{ps}\sim \Delta x ^N \sim 1/N^N
$$
and for the spectral element error bound \cite[p. 273]{DFM2002}, that 
$$
\epsilon_{se}\sim h^{\min(p,s)} \ {p}^{-s}
$$
where $h\sim1/E$ is the uniform element length, and $s$ is the smoothness of the exact solution.
It is clear that in practice $s\le p$, since the derivatives for $s > p$ cannot be computed.
We thus choose $s=p$ so that the function 
is sufficiently smooth to allow spectral convergence.

Equating the logarithmic errors 
immediately shows the relationship between $N$, $E$ and $p$, namely:
$$
N\log N \sim \ p \log (p\ E).
$$

The above scaling argument allows for choosing a range of values of  polynomial order under simple assumptions.
Let us say the Reynolds number is doubled from a well-adjusted run; roughly speaking, we need to  double the 
resolution of the pseudo-spectral code from $N$ to $2N$  grid points per linear dimension. Let us also assume 
that we double  the number of elements in the spectral element code. Then, under the  reasonable assumption of 
large N, we find that the polynomial order  needs to be doubled as well. Though empirical, this criterion does  
indicate that the polynomial order needs to increase with Reynolds  number. It should be noted that, 
whereas the pseudo-spectral code, in  the preceding example of \fig{fig_jmax_128_p}, uses a truncation of order 
$128$, the equivalent spectral element code uses polynomials of order  $5$, substantially smaller, in order 
to achieve comparable accuracy.

\begin{figure}
\begin{center}\includegraphics[width=8.3cm]{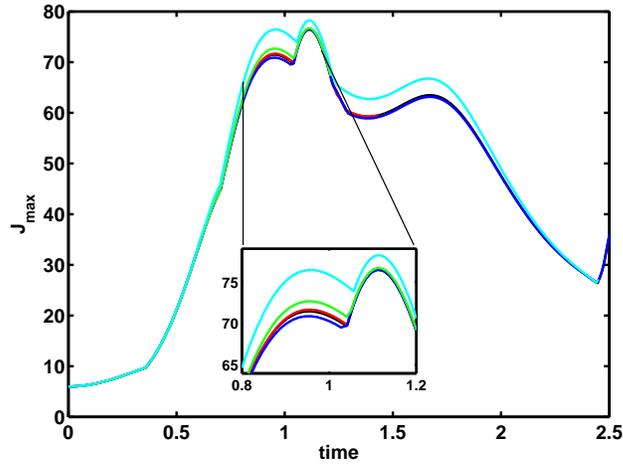}
\end{center}
\caption{Maximum of current $J_{\rm max}$ as a function of time for the second series of $N_{\rm eq} \approx 256 $
runs in Table \ref{tbl_runs}. The black curve
is the pseudo--spectral case; red is the $p=8$ run; dark blue is the $p=6$ run; green is the
$p=4$ case; cyan is the $p=3$ run. Note that this time the $p=8$ and pseudo--spectral results are
nearly coincident. } 
\label{fig_jmax_256_p}
\end{figure}

\begin{figure}
\begin{center} \includegraphics[width=8.3cm]{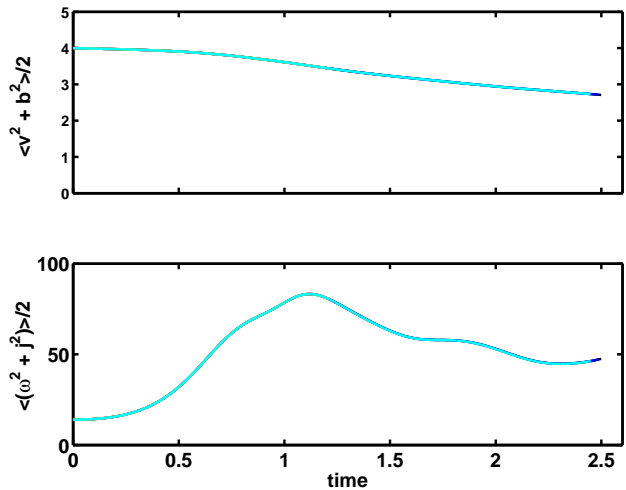} \end{center}
\caption{Plots of total energy ({\it top}) and enstrophy ({\it bottom}) for the  $N_{\rm eq} \approx 256 $ 
runs in Table \ref{tbl_runs}. The color scheme is the same as in \fig{fig_jmax_256_p}, but the
curves lie on top of one another. }
\label{fig_bal_p}
\end{figure}

As a specific example, an examination of \fig{fig_jmax_128_p} indicates  that, with $E=26$, setting $p=5$ leads to a 
satisfactory computation of $J_{\rm max}$. Let us now double the resolution of the  pseudo-spectral 
code to $N^{\prime}=256$ and take
$E^{\prime}=52$. As can be seen in \fig{fig_jmax_256_p}, it is indeed  the case that $p\ge8$ gives an accurate 
representation of the dynamics  of the flow at that enhanced Reynolds number. According to the scaling relationship,
if we were to retain $p=5$, we would need $E\approx10000$ in order to reach the same level of accuracy at the higher
$\Rn$.
For completeness, we show in figure \ref{fig_bal_p} the 
corresponding ${\cal L}_2$ norms for the total energy (top) and the total generalized enstrophy 
$<\omega^2+j^2>$ for the $N_{\rm eq} \approx 256$ cases; the different runs cannot be discerned.


%
%
%
%
%
%
%
%
%
%

\section{Discussion and Conclusion} 
\label{sec_conclusion}

We have presented an explicit spectral element method for solving the equations of 
incompressible magnetohydrodynamics. The method is developed within the context of
an existing spectral element code (\gaspar) that provides many of the spectral element
operators required of the MHD algorithm, and that also offers an adaptive, nonconforming
mesh algorithm. The new operators that arise in the explicit MHD treatment, have been
defined.  Here, we have described tests that compare the numerical results with 
analytic solutions and establish that the method achieves spectral convergence 
in the case of conforming elements (for preliminary tests on a static reconnection 
problem, see \cite{NgDPP2006}). We have then applied the method to a challenging problem in 
the literature, the so-called Orszag-Tang flow, which allows for 
magnetic reconnection and the development of current sheets. 
The OT runs are compared with well-tested pseudo--spectral solutions as a baseline, and found to
agree well. We find that the quadratic $\Ltwo$ 
diagnostic quantities are insensitive to variations in polynomial degree, while the 
sup--norm quantities, such as the maximum current density, are not accurate at low order truncations.
Because such sup-norm quantities are the foundation for criteria of the development (or not) of singularities 
in Navier-Stokes and MHD flows \cite{singu_NS,singu_MHD}, it may be of some importance to be able 
to solve for them accurately.

It is worth pointing out that the spectral element method described in this paper is more 
costly in terms of
computational time than the pseudo--spectral method with which we compare our results, the latter 
being optimal for periodic boundary conditions. This is despite
the fact that the spectral element method requires only nearest--neighbor communication, while the
pseudo--spectral method requires all--to--all exchange of data in the fast Fourier transform algorithm. 
This performance issue can be traced directly to the
solutions of the pseudo--Poisson equation \eq{eq_pseudopoisson} that are required 
in order to maintain the divergence constraints. The pseudo--Laplacian operator is
known to be ill--conditioned, primarily because of a one-dimensional null space. Indeed,
for the $N_{\rm eq}=512^2$ equivalent runs, we see typical iteration counts for each RK stage of $\sim 250$, which further
increase--albeit reasonably slowly--in the OT runs as the enstrophy maximum is approached; we see 
no significant reduction in the PCG iteration count when we attempt to remove this
null--space, but more work is needed in this area.
[It should be noted that the scaling of the spectral element code on multiprocessors is very good, but we 
refer here to single processor performance.] 
Furthermore, when dealing with incompressible MHD fluids at low magnetic Prandtl number, as encountered 
in the laboratory and in the liquid core of the Earth, there is a need for accurate simulations of the generation 
of magnetic fields in turbulent flows with complex boundaries, in conjunction with several ongoing experiments
(see e.g., \cite{Karlsruhe,Riga,Madison,pinton}). Spectral element codes, which encompass easily a variety of 
boundary conditions and geometries may be useful in this context.

There are a number of ways to speed up the spectral element solutions. One is to implement a
more sophisticated preconditioner, and we are making progress in this regard that will be
reported on elsewhere. Another is to relax
the degree to which the divergence constraints are maintained, by increasing the PCG convergence
tolerance to a less stringent value. Still another, as alluded to in \Sec{sec_ot}, is to find refinement
criteria that optimize the number of degrees of freedom for a desired quadrature or truncation error.
We have seen that the adaptive mesh code can provide a substantial savings in work because
the number of degrees of freedom, as shown in several instances (\eg \cite{rosenberg2006}), can be 
reduced. 
The code is presently being extended to include compressibility in view 
of the many MHD applications we have in mind in the astrophysical context (solar wind, magnetosphere, 
solar convection zone and corona), in which case the performance problems associated with 
this divergence-free constraint should be alleviated. 




\ack
We acknowledge helpful discussions with Amik St.Cyr and Aim\'e Fournier, 
and we also thank the latter for providing some of the analysis and
visualization software used in this work. 
Computer time was provided by NCAR. The 
NSF grant CMG-0327888 at NCAR supported this work in part and is gratefully 
acknowledged. 

\end{document}